\documentclass[aps,prl,showpacs,twocolumn,superscriptaddress]{revtex4}
\usepackage{graphicx}
\usepackage{amssymb}
\usepackage{epsfig}
\usepackage{graphicx}
\usepackage{amsmath}
\usepackage{array,color}

\begin{document}

\title{Motion of solitons in one-dimensional spin-orbit-coupled
Bose-Einstein condensates}
\author{Lin Wen}
\affiliation{Department of Physics, Chongqing Normal University, Chongqing, 401331, China}
\affiliation{Department of Physics, Capital Normal University, Beijing 100048, China}
\author{Q. Sun}
\email{QingSun@cnu.edu.cn}
\affiliation{Department of Physics, Capital Normal University, Beijing 100048, China}
\author{Yu Chen}
\affiliation{Department of Physics, Capital Normal University, Beijing 100048, China}
\author{Deng-Shan Wang}
\affiliation{School of Applied Science, Beijing Information Science and Technology
University, Beijing 100192, China}
\author{J. Hu}
\affiliation{Department of Physics, Capital Normal University, Beijing 100048, China}
\author{H. Chen}
\affiliation{Institute of Physics, Chinese Academy of Sciences, Beijing 100190, China}
\author{W.-M. Liu}
\affiliation{Institute of Physics, Chinese Academy of Sciences, Beijing 100190, China}
\author{G. Juzeli\={u}nas}
\affiliation{Institute of Theoretical Physics and Astronomy, Vilnius University, Saul\.{e}%
tekio Ave. 3, LT-10222 Vilnius, Lithuania}
\author{Boris A. Malomed}
\affiliation{Department of Physical Electronics, School of Electrical Engineering,Faculty
of Engineering, Tel Aviv University, Tel Aviv 69978, Israel}
\affiliation{Laboratory of Nonlinear-Optical Informatics, ITMO University, St. Petersburg
197101, Russia}
\author{An-Chun Ji}
\email{anchun.ji@cnu.edu.cn}
\affiliation{Department of Physics, Capital Normal University, Beijing 100048, China}

\begin{abstract}
Solitons play a fundamental role in dynamics of nonlinear excitations. Here
we explore the motion of solitons in one-dimensional uniform Bose-Einstein
condensates subjected to a spin-orbit coupling (SOC). We demonstrate that
the spin dynamics of solitons is governed by a nonlinear Bloch equation. The
spin dynamics influences the orbital motion of the solitons leading to  the
spin-orbit effects in the dynamics of the macroscopic quantum objects
(mean-field solitons). The latter perform oscillations with a frequency
determined by the SOC, Raman coupling, and intrinsic nonlinearity. These
findings reveal unique features of solitons affected by the SOC, which is
confirmed by analytical considerations and numerical simulations of the
underlying Gross-Pitaevskii equations.
\end{abstract}

\maketitle

Solitons, which are generally realized as self-supported solitary waves, are
among the most fundamental objects in the nonlinear science. With the
realization of Bose-Einstein condensates (BEC), matter-wave solitons have
drawn enormous interest \cite{Randy,Brazh,Fatkh,Morsch,fsm,soliton1,Dim,RMP2,vs}. In the experiments,
both bright and dark solitons have been successively created in atomic BEC
\cite{Randy,bs1,soliton2,soliton3,soliton4,soliton5}. On the other hand, the
successful realization of the artificial spin-orbit coupling (SOC) in binary
BEC \cite{soc1,soc2,soc3,soc4} has stimulated intensive studies on novel
SOC-induced effects \cite{Zhaih,TLHo,Lyou,CWZhang1,YLi,SZhang,HHu,CWZhang2,ZQYu,CWZhang3,CWZhang4,WYi,Galitski2013,Goldman2014,Zhai2015,Su2016}. In particular, a variety of solitons species,
such as stripe modes, 2D composite solitons, and half-vortex gap soliton have been predicted in the
condensates combining the SOC, which is a linear interaction, and the
intrinsic collisional nonlinearity \cite{ss1,ss4,ss3,ss2,ss5,ss6,ss7,ss8,ss9,bdd,sgs,2cs1,Luca,2cs2,hvgs,Petra,YCZhang,Fialko}.

Soliton dynamics has been the subject of many studies carried out in various
settings, including BEC \cite{Randy,Brazh,Fatkh,Morsch,fsm,soliton1,Dim,RMP2,vs}, nonlinear optics \cite%
{RMP1,KivPel,KA,bullets,DesTorKiv,Dum,PT}, and others.{\LARGE \ }In
particular, it has been shown that harmonic trapping potentials induce
motion of solitons in quasi-one-dimensional (1D) BEC \cite%
{d10,d9,d7,d11,HPu,d12}. Due to the particle-like nature, the soliton
dynamics differs essentially from the dipole mode of the non-interacting
condensate loaded into the same potential. In particular, the oscillation
frequency of trapped dark solitons differs by factor $1/\sqrt{2}$ from the
trap frequency \cite{d10}. Here, we address the soliton dynamics in 1D BEC
under the action of the Raman-induced SOC \cite{soc1,soc2}. Since the Raman
transition can flip the spin along with inducing a finite momentum transfer,
the evolution of the spin degree of freedom may be coupled to the spatial
motion of solitons. This effect, if it can be made conspicuous enough, may
be considered as the SOC at the level of the motion of a macroscopic quantum
body, the matter-wave soliton. In this connection, it is relevant to mention
a recent result showing that the SOC can induce anharmonic properties beyond
the effective-mass approximation in collective dipole oscillations \cite%
{soc2,ChenZ,LiY}. Yet the macroscopic SOC effects in the motion of solitons
have not been explored before, to the best of our knowledge.

In this work we investigate the soliton dynamics in 1D uniform BEC
influenced by the SOC. We find that an interplay of the SOC, Raman coupling,
and nonlinearity induces precession of the soliton's spin $\mathbf{S}$ under
the action of an effective magnetic field, which is governed by a nonlinear
Bloch equation (\ref{Bloch}). In turn, the spin precession couples to the
orbital motion of the soliton via feedback onto its center-of-mass momentum,
as shown below by equation of motion (\ref{xi}) for the center-of-mass
coordinate, $\left\langle z\right\rangle $. Thus, Eqs. (\ref{Bloch}) and (%
\ref{xi}) directly demonstrate the effects of the SOC on the 1D motion of
the macroscopic quantum object.

In the presence of SOC, the dynamics of the quasi-1D BEC, elongated in the
direction of $z$, is modelled by the mean-field Gross-Pitaevskii (GP)
equation: 
\begin{equation}
i\partial _{t}\left( \!%
\begin{array}{c}
\psi _{\uparrow } \\
\psi _{\downarrow }%
\end{array}%
\!\right) \!=\!\hat{h}_{0}\left( \!%
\begin{array}{c}
\psi _{\uparrow } \\
\psi _{\downarrow }%
\end{array}%
\!\right) \!+\!\left(
\begin{array}{c}
g_{\uparrow \uparrow }|\psi _{\uparrow }|^{2},g_{\uparrow \downarrow }|\psi
_{\downarrow }|^{2} \\
g_{\downarrow \uparrow }|\psi _{\uparrow }|^{2},g_{\downarrow \downarrow
}|\psi _{\downarrow }|^{2}%
\end{array}%
\right) \!\left( \!%
\begin{array}{c}
\psi _{\uparrow } \\
\psi _{\downarrow }%
\end{array}%
\!\right) ,  \label{GP}
\end{equation}%
where $\psi _{\sigma }$ are the pseudo-spin components of the BEC
macroscopic wave function, with $\sigma =\uparrow ,\downarrow $ labelling
the spin states. These can represent, for instance, the hyperfine states $%
|1,-1\rangle $ and $|1,0\rangle $ of $^{87}$Rb atoms \cite{soc1}. Here
\begin{equation}
\hat{h}_{0}=-\frac{1}{2}\partial _{z}^{2}+V(z)+i\lambda \partial _{z}\sigma
_{z}+\Omega \sigma _{x}+\delta \sigma _{z}  \label{hh0}
\end{equation}%
is a single-particle Hamiltonian which includes the Raman-induced SOC
characterized by a strength $\lambda $, with $\Omega $ and $\delta $
describing, respectively, the frequencies of the Raman coupling and the
Zeeman detuning. Here also $V(z)=\gamma ^{2}z^{2}/2$ is an effective 1D
harmonic trap potential, and $\gamma \equiv \omega _{z}/\omega _{\bot }$ is
the trap's aspect ratio, with $\omega _{z}$ and $\omega _{\bot }$ being the
trapping frequencies along the longitudinal and transverse directions,
respectively. The frequencies and lengths are measured in units $\omega
_{\bot }$ and $a_{\bot }=\sqrt{\hbar /m\omega _{\bot }}$, respectively, and,
as mentioned above, $\lambda =k_{L}a_{\bot }$ represents the SOC\ strength,
with $k_{L}$ being the momentum transfer. Note that, as the strengths of the
inter- and intra-species atomic interactions are very close in the
experiment, it is reasonable to assume $SU(2)$-symmetric spin interactions,
with all components $g_{\sigma \sigma ^{\prime }}$ taking a single value, $g$%
. To focus on the SOC effects on the dynamics of solitons, we first consider
the free space, while the external trap will be discussed afterwards.

For $\lambda =\Omega =0$, the system reduces to a normal binary BEC without
the SOC. In this case, Eq. (\ref{GP}) is known as the integrable Manakov's
system which gives rise to well-known exact soliton solutions \cite{manakov}%
. In particular,bright-bright (BB) solitons are $\psi _{\sigma }=\left( \eta
\epsilon _{\sigma }/\sqrt{-g}\right) $sech$\left( \eta z\right) \exp \left(
i\eta ^{2}t/2\right) $ for the attractive sign of the nonlinearity, $g<0$,
where $\eta ^{-1}$ is the soliton's width and $\epsilon _{\sigma }$
satisfies the normalized condition, $|\epsilon _{\uparrow }|^{2}+|\epsilon
_{\downarrow }|^{2}=-g/(2\eta )$. We use such exact soliton solutions as an
initially prepared wave function, and then study the soliton dynamics as the
SOC is switched on. Note that Eq. (\ref{hh0}) is an effective
single-particle Hamiltonian in the frame transformed via the local
pseudo-spin rotation by angle $\vartheta =2\lambda z$ about the $z$ axis
\cite{soc1,soc2}. The transformation adds opposite phase factors, $e^{\pm
i\lambda z}$, to the two components of the input waveforms.

In the general case, the GP system (\ref{GP}) is no longer integrable.
Therefore we employ a variational approximation to investigate the soliton
dynamics \cite{Progress,fsm}, based on Lagrangian $L(t)=\int_{-\infty
}^{+\infty }\{(i/2)\sum_{\sigma =\uparrow ,\downarrow }\left[ \psi _{\sigma
}^{\ast }\left( \psi _{\sigma }\right) _{t}-\psi _{\sigma }\left( \psi
_{\sigma }^{\ast }\right) _{t}\right] -\mathcal{H}\}dz$, where $\mathcal{H}$
is the Hamiltonian density of the system. We first consider the attractive
nonlinearity, with $g<0$. In this case, we introduce the following
variational Ansatz for BB solitons, with the total norm fixed to be $1$:
\begin{equation}
\left(
\begin{array}{c}
\psi _{\uparrow } \\
\psi _{\downarrow }%
\end{array}%
\right) =\sqrt{\frac{\eta }{2}}\left(
\begin{array}{c}
\left( \sin \theta \right) \>\text{sech}(\eta z+\xi _{\uparrow })e^{i\left(
k_{\uparrow }z+\varphi _{\uparrow }\right) } \\
\left( \cos \theta \right) \>\text{sech}(\eta z+\xi _{\downarrow
})e^{i\left( k_{\downarrow }z+\varphi _{\downarrow }\right) }%
\end{array}%
\right) \,,  \label{ansatz}
\end{equation}%
where $\theta $, $\eta $, $\xi _{\sigma }$, $k_{\sigma }$, $\varphi _{\sigma
}$ are time-dependent variational parameters. Here $\theta $ determines the
population imbalance between the pseudo-spin components, $\eta ^{-1}$
defines their common width, $k_{\sigma }$ is the wavenumber, and $\varphi
_{\sigma }$ the phase. For the $SU(2)$ atomic interactions, two-component
solitons favor the mixed phase \cite{phasemixing}. Hence the positions of
spin up and down solitons will overlap, $\xi _{\uparrow }=\xi _{\downarrow
}=\xi $, as confirmed by the numerical simulations below.

Inserting the Ansatz (\ref{ansatz}) into the Lagrangian and performing the
integration, we obtain
\begin{eqnarray}
L\left( t\right) &=&\frac{\xi }{\eta }\frac{dk_{+}}{dt}-\frac{\xi }{\eta }%
\cos \left( 2\theta \right) \frac{dk_{-}}{dt}-\frac{d\varphi _{+}}{dt}+\cos
\left( 2\theta \right) \frac{d\varphi _{-}}{dt}  \notag \\
&-&\frac{1}{2}\left[ k_{+}^{2}-2k_{+}k_{-}\cos \left( 2\theta \right)
+k_{-}^{2}\right] -\frac{1}{6}\eta ^{2}-\frac{1}{6}g\eta  \notag \\
&-&\frac{\Omega \pi k_{-}\sin (2\theta )\cos \left( 2\varphi _{-}-2k_{-}\xi
/\eta \right) }{\eta \sinh \left( \pi k_{-}/\eta \right) }  \notag \\
&+&\delta \cos \left( 2\theta \right) -\lambda \left[ k_{+}\cos \left(
2\theta \right) -k_{-}\right] \,,  \label{L}
\end{eqnarray}%
where $k_{\pm }\equiv \left( 1/2\right) (k_{\uparrow }\pm k_{\downarrow })$
and $\varphi _{\pm }\equiv \left( 1/2\right) (\varphi _{\uparrow }\pm
\varphi _{\downarrow })$. The evolution of the variational parameters is
governed by the corresponding Euler-Lagrangian (EL) equations, see the
Supplementary material for details.
\begin{figure}[tbp]
\includegraphics[width= 0.46\textwidth]{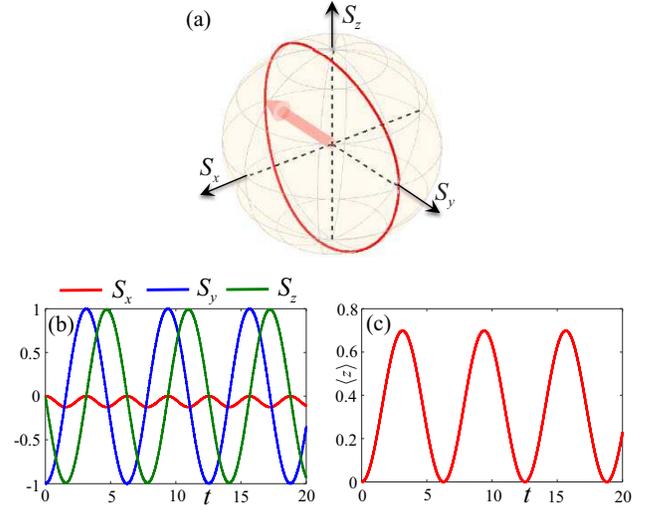}
\caption{(Color online). The track of the spin density on the Bloch sphere
(a), and the corresponding evolution of the spin components (b) and
center-of-mass coordinate (c) of the soliton for the initially balanced
state, with $\protect\theta (t=0)=\protect\pi /{4}$, and initial phase
difference $\protect\varphi _{-}(t=0)=\protect\pi /4$ [see Eq. (\protect\ref%
{ansatz})]. Other parameters are $\Omega =0.5$, $\protect\lambda =0.5\protect%
\sqrt{\Omega }$, $\protect\delta =0$, and $g=-10$.}
\label{fig1}
\end{figure}
Note that the EL equations produce simple results, $\eta \approx -g/2$ and $%
k_{-}\approx \lambda $, in the case of a weak SOC, $\pi \lambda <<\eta $.
Indeed, in the absence of SOC, the relation $\eta =-g/2$ holds for the
normalized wave function, indicating that the width of the solitons is
determined by the nonlinearity, and $k_{-}$ remains equal to the initial
relative momentum $\lambda $ between the components of the soliton. Thus, we
arrive at a reduced system of the EL equations, in which $\eta $ and $k_{-}$
are considered as frozen quantities:
\begin{eqnarray}  \label{ELE}
\dot{k}_{+} &=&2\lambda \tilde{\Omega}\sin ({2\theta })\sin {\phi }\,,
\label{EL11} \\
\dot{\phi} &=&-2\tilde{\Omega}\cot (2\theta )\cos \phi +2\lambda
k_{+}-2\delta \,,  \label{EL22} \\
\dot{\theta} &=&-\tilde{\Omega}\sin \phi \,,  \label{EL33} \\
\dot{\langle z\rangle } &=&k_{+}\,.  \label{EL44}
\end{eqnarray}%
Here, $\left\langle z\right\rangle =\int_{-\infty }^{+\infty }z(|\psi
_{\uparrow }|^{2}+|\psi _{\downarrow }|^{2})dz\equiv -\xi /\eta $ is the
center-of-mass coordinate, $\tilde{\Omega}\equiv \Omega \pi \left( \lambda
/\eta \right) \sinh \left( \pi \lambda /\eta \right) $, and $\phi \equiv
2\varphi _{-}+2k_{-}\langle z\rangle $ is the phase difference between the
two components of the soliton. Thus, Eqs. (\ref{EL11})-(\ref{EL44}) account
for the nonlinear coupling of the center-of-mass momentum $k_{+}$, phase
difference $\phi $, population imbalance $\theta $, and center-of-mass
coordinate ${\langle z\rangle }$.

We introduce a normalized complex-valued spinor, $\mathbf{\chi }=(\chi
_{\uparrow },\chi _{\downarrow })$ describing the two-component wave
function $\psi _{\sigma }=\sqrt{\rho \left( z,t\right) }\chi _{\sigma }$,
where $\rho \equiv |\psi _{\uparrow }|^{2}+|\psi _{\downarrow }|^{2}$ is the
total density, with $\left\vert \chi _{\uparrow }\right\vert ^{2}+\left\vert
\chi _{\downarrow }\right\vert ^{2}=1$. Furthermore we define the spin
density $\mathbf{S}=\mathbf{\chi }^{T}\mathbf{\sigma \chi }$, where $\mathbf{%
\sigma \equiv }\left\{ \sigma _{x},\sigma _{y},\sigma _{z}\right\} $ is a
vector set of the Pauli matrices, we have $\{S_{x},S_{y},S_{z}\}=\{\sin
(2\theta )\cos (2\lambda z+2\varphi _{-}),$$-\sin (2\theta )\sin (2\lambda
z+2\varphi _{-}),-\cos (2\theta )\}\equiv \{\sin (2\theta )\cos \phi ,-\sin
(2\theta )\sin \phi ,-\cos (2\theta )\}$ for the Ansatz given by Eq. (\ref%
{ansatz}). In terms of the soliton spin components, Eqs. (\ref{EL11})-(\ref%
{EL33}) are then written as
\begin{eqnarray}
\dot{S}_{x} &=&2(\lambda c_{1}-\delta )S_{y}-2\lambda ^{2}S_{z}S_{y}\,,
\label{sx} \\
\dot{S}_{y} &=&-2\tilde{\Omega}S_{z}-2(\lambda c_{1}-\delta )S_{x}+2\lambda
^{2}S_{z}S_{x}\,,  \label{sy} \\
\dot{S}_{z} &=&2\tilde{\Omega}S_{y}\,,  \label{sz}
\end{eqnarray}%
where $c_{1}\equiv \lambda S_{z,0}$, with $S_{\alpha ,0}$ ($\alpha =x,y,z$)
being the initial values of the components. These equations of motion for the
soliton's spin can be rewritten as
\begin{equation}
\dot{\mathbf{S}}=\mathbf{S}\times \mathbf{B}_{\mathrm{eff}},\ \mathbf{B}_{%
\mathrm{eff}}=\left\{ 2\tilde{\Omega},0,2\lambda ^{2}S_{z}-2(\lambda
c_{1}-\delta )\right\} .  \label{Bloch}
\end{equation}%
This represents the Bloch equation for the spin precession under the action
of the effective magnetic field, $\mathbf{B}_{\mathrm{eff}}$. The
macroscopic SOC for the soliton as a quantum body is determined by the
effect of evolution of the spin on the soliton's longitudinal momentum,
resulting in the coupled nonlinear dynamics of the soliton's spin and
position. At the first glance, nonlinear terms in Eqs. (\ref{sx})-(\ref{sz})
arise essentially from the SOC\ strength, $\lambda $. However, the atomic
interactions also play a fundamental role, as in the no-interaction limit, $%
\tilde{\Omega}=0$, these equations reduce to the linear Bloch precession
under a fixed effective magnetic field.

To tackle solutions of the nonlinear Bloch equation, we first integrate Eq. (%
\ref{sx}), dividing it by Eq. (\ref{sz}). This yields $S_{x}=c_{2}-\left(
\lambda ^{2}/2\right) \tilde{\Omega}^{-1}S_{z}^{2}+(\lambda c_{1}-\delta )%
\tilde{\Omega}^{-1}S_{z}$, where $c_{2}\equiv S_{x,0}+\left( \delta
-c_{1}\lambda \right) \tilde{\Omega}^{-1}S_{z,0}+\left( \lambda
^{2}/2\right) \tilde{\Omega}^{-1}S_{z,0}^{2}$ is a constant determined by
the initial conditions. Next, we focus on the case of $\delta =\lambda c_{1}$%
, which implies a particular relation between the strengths of the Zeeman
splitting and SOC, making the analysis more explicit. By differentiating Eq.
(\ref{sz}), we then arrive at a standard equation of anharmonic oscillations
for the single spin component, $S_{z}$,
\begin{equation}
\frac{d^{2}S_{z}}{dt^{2}}+\Xi S_{z}+2\lambda ^{4}S_{z}^{3}=0\,,
\label{nonlinear oscillator}
\end{equation}%
where $\Xi \equiv 4\tilde{\Omega}(\tilde{\Omega}-c_{2}\lambda ^{2})$ may be
positive or negative. Equation (\ref{nonlinear oscillator}) has a usual
solution \cite{Jacobi}
\begin{equation}
S_{z}(t)=\frac{\sqrt{1-\Xi \tau ^{2}}}{2\lambda ^{2}\tau }\mathrm{cn}\left(
t/\tau ,k\right) ,~k^{2}=\frac{1}{2}\left( 1-\Xi \tau ^{2}\right) ,
\label{cn}
\end{equation}%
where $\mathrm{cn}$ is the Jacobi's cosine with modulus $k$, and $\tau $ is
an arbitrary parameter taking values $\tau <1/\sqrt{|\Xi |}$. In the case of
$\Xi >0$, the linearized version of Eq. (\ref{nonlinear oscillator}), which
corresponds to $\tau \rightarrow 1/\sqrt{|\Xi |}$ in Eq. (\ref{cn}), gives
rise to free Rabi oscillations with frequency $\sqrt{\Xi }$. In the general
case, the frequency given by solution (\ref{cn}), $\omega _{\mathrm{osc}%
}=\pi /(2\tau K(k))$ exceeds $\sqrt{\Xi }$ due to the nonlinear
shift, where $K(k)$ is the complete elliptic integral. Note also
that the nonlinearity may give rise to oscillations in the case of
$\Xi <0$, when the free Rabi oscillations are impossible.
\begin{figure}[tbp]
\includegraphics[width= 0.48\textwidth]{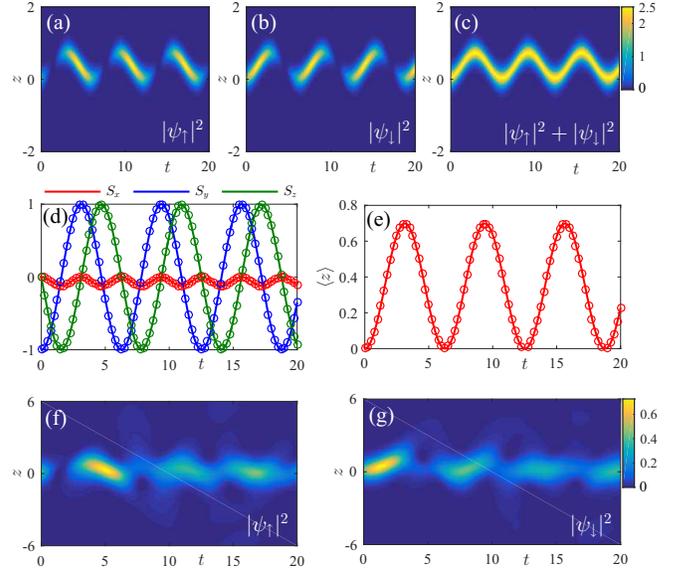}
\caption{(Color online). (a)-(c) The evolution of the density in the two
components of the soliton produced by the GP simulations for the initially
balanced state with $\protect\theta (t=0)=\protect\pi /{4}$ and $\protect%
\varphi _{-}(t=0)=\protect\pi /4$. The parameters are $\Omega =0.5$, $%
\protect\lambda =0.5\protect\sqrt{\Omega }$, $\protect\delta =0$, and $g=-10$%
. The spin dynamics and center-of-mass motion of the soliton, generated by
these simulations (circles), and by the variational approximation based on
Eqs. (\protect\ref{sx})-(\protect\ref{sz}) and (\protect\ref{xi}) (solid
lines) are depicted in panels (d)-(e). Panels (f)and(g) display decay decay
of the soliton in the case of weaker weaker atomic interaction, $g=-3$.}
\label{fig2}
\end{figure}

Further, in the spin representation, Eq. (\ref{EL11}) can be written as $%
\dot{k}_{+}=-2\lambda \tilde{\Omega}S_{y}$, which accounts for the effect of
the evolution of the spin on the center-of-mass momentum. This leads to the
following equation of motion for the center-of-mass coordinate:%
\begin{equation}
\frac{d^{2}\left\langle z\right\rangle }{dt^{2}}=-\lambda \tilde{\Omega}%
S_{y}\,.  \label{xi}
\end{equation}%
In other words, if we consider the soliton as a macroscopic quantum body
carrying the intrinsic angular momentum, Eq. (\ref{xi}) represents the
driving force, exerted by the intrinsic momentum and acting on the linear
momentum, which is literally the macroscopic SOC. We stress that the motion
of the soliton differs from the collective dipole oscillations of BEC in a
harmonic-potential trap under the action of the SOC, which is a wave effect.
On the other hand, here we consider an effectively mechanical motion, which
is produced by the interplay of the nonlinear self-trapping and SOC.

To illustrate the soliton dynamics in detail, we display, in Fig.\ref{fig1},
numerical solutions of Eqs. (\ref{sx})-(\ref{sz}) and (\ref{xi}), with
initially balanced populations in the two components, which corresponds to $%
\theta (t=0)=\pi /4$ and $\varphi _{-}(t=0)=\pi /4$. First, in Fig. \ref%
{fig1}(a) we show that the soliton spin moves along a closed orbit on the
Bloch sphere. Accordingly, perfect periodic oscillations of the spin can be
identified in Fig. \ref{fig1}(b), and perfectly periodic linear motion of
the soliton's central coordinate is seen in Fig. \ref{fig1}(c).

To test these findings obtained in the variational (i.e., effectively
mechanical) approximation, we numerically solved the GP system (\ref{GP})
with the input in the form of BB solitons, $\psi _{\uparrow }=\sqrt{\eta /2}%
\left( \sin \theta _{0}\right) \>\text{sech}(\eta z)e^{i(\lambda z+\varphi
_{\uparrow ,0})}$ and $\psi _{\downarrow }=\sqrt{\eta /2}\left( \cos \theta
_{0}\right) \>\text{sech}(\eta z)e^{i(-\lambda z+\varphi _{\downarrow ,0})}$%
. In Figs. \ref{fig2} (a)-(c), the density of the soliton's components
exhibits remarkable periodic oscillations. Note that, although the initial
momenta of the two components are opposite, the soliton does not split. In
Fig. \ref{fig2} (d)-(e), we show that the direct GP simulations agree very
well with the variational (mechanical) approximation. For weaker atomic
interactions, the simulations show that solitons decay under the action of
SOC for $\pi \lambda /\eta >(\pi \lambda /\eta )_{c}\approx 0.4$, as shown
in Fig. \ref{fig2} (f)-(g). We have also explored the evolution initiated by
other inputs, such as, e.g., the initially polarized one with $\theta(t=0)
=\pi /{2}$ and $\varphi _{-}(t=0)=0$, and obtained  a similar dynamical
behavior (see the Supplementary material for details).

We stress that these results are relevant for $\lambda \leq \sqrt{\Omega }$,
where the spectrum of the single-particle Hamiltonian with the SOC terms has
one minimum, and the evolution of the initial BB solitons is stable. At $%
\lambda >\sqrt{\Omega }$, the single-particle spectrum develops two
degenerate minima, and the initially prepared solitons gradually decay under
the action of the SOC, as demonstrated by the GP simulations.
\begin{figure}[t]
\includegraphics[ width= 0.48\textwidth]{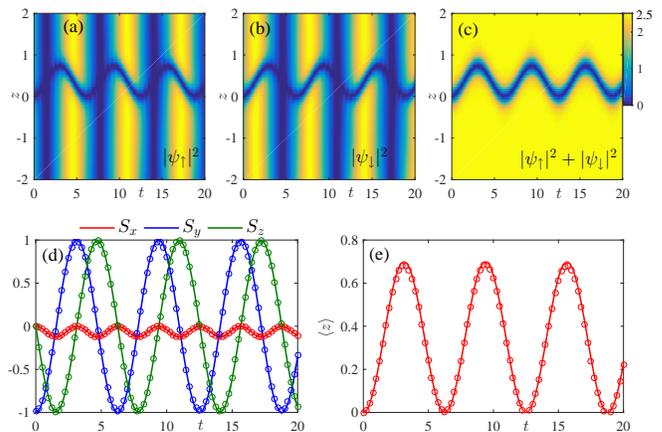}
\caption{(Color online). (a)-(c) The same as in Fig. \protect\ref{fig2}, but
for the dark soliton, in the case of $g=10$.}
\label{fig3}
\end{figure}

Now, we proceed to the case of the repulsive BEC nonlinearity, which is more
relevant to current experiments with SOC condensates ~\cite%
{soc1,soc2,soc3,soc4}. In this case, we introduce the following variational
Ansatz for dark-dark (DD) solitons:
\begin{equation}
\left(
\begin{array}{c}
\psi _{\uparrow } \\
\psi _{\downarrow }%
\end{array}%
\right) =\sqrt{\frac{\eta }{2}}\left(
\begin{array}{c}
\left( \sin \theta \right) \tanh (\eta z+\xi )e^{i\left( k_{\uparrow
}z+\varphi _{\uparrow }\right) } \\
\left( \cos \theta \right) \tanh (\eta z+\xi )e^{i\left( k_{\downarrow
}z+\varphi _{\downarrow }\right) }%
\end{array}%
\right) \,.  \label{ansatz1}
\end{equation}%
Inserting the Ansatz (\ref{ansatz1}) into the Lagrangian, one needs to
renormalize the integrals to exclude divergent contributions of the
nonvanishing background \cite{r1,r2,r3,r4,r5}. The analysis
yields the same EL equations as Eqs. (\ref{EL22})-(\ref{EL44}), but with $%
\eta =g/2$ for repulsive $g>0$. We have also performed the respective direct
simulations of the GP system with the initial conditions corresponding to
the DD solitons, $\psi _{\uparrow }=\sqrt{\eta /2}\left( \sin \theta
_{0}\right) \tanh (\eta z)e^{i(\lambda z+\varphi _{\uparrow ,0})}$ and $\psi
_{\downarrow }=\sqrt{\eta /2}\left( \cos \theta _{0}\right) \>\tanh (\eta
z)e^{i(-\lambda z+\varphi _{\downarrow ,0})}$. The results are depicted in
Fig. \ref{fig3}, where the DD soliton displays oscillations similar \ to
those reported above for the BB configuration. Note that the two components
of the background alternately disappear and revive, as shown in Fig. \ref%
{fig3} (a)-(b).

In experiments, the condensate is usually trapped in a harmonic-oscillator
potential, $V(z)=\gamma ^{2}z^{2}/2$, which affect the motion of solitons
\cite{d10,d9,d7,d11,HPu,d12}. In this case, the EL equations for variational
parameters $k_{+}$ and $\xi $ are modified as $\dot{k}_{+}=-2\lambda \tilde{%
\Omega}S_{y}+\left( \gamma ^{2}/\eta \right) \xi $ and $\dot{\xi}=-{k}%
_{+}\eta $ with $\eta $ satisfying condition $4\eta ^{4}+2g\eta ^{3}=\pi
^{2}\gamma ^{2}$, cf. Eqs. (\ref{EL11}) and (\ref{EL44}). For the quasi-1D
cigar-shaped BEC in our case, $\omega_{z}\ll\omega_{\bot}$, we have $\gamma
\sim 0$, and the effects of trapping potential may be neglected. On the
other hand, for strong trap potentials, the soliton dynamics becomes quite
complex, due to the strong coupling between the spatial inhomogeneity and
SOC. This issue will be considered elsewhere.

Finally, we discuss some related experimental issues. So far, the
Raman-induced SOC has been realized for the BEC in the $^{87}$Rb gas with
repulsive atomic interactions. The corresponding ratios of the scattering
lengths, $a_{\uparrow \uparrow }:a_{\uparrow \downarrow }:a_{\downarrow
\downarrow }=1:1:1.005$ \cite{soc1}, corroborate our assumption of equal
interaction strengths. In this case, the DD solitons can be created by means
of the phase- and density-engineering techniques \cite{soliton3}. We
consider an elongated condensate made of $\sim 10^{4}$ atoms under trapping
frequencies $\omega _{\bot }=2\pi \times 2.7$ kHz and $\omega _{z}=2\pi
\times 26$ Hz. The recoil momentum $k_{L}$ can be adjusted by the angle
between the two incident Raman beams. For example, for the $804.1$ nm Raman
lasers intersecting at angle $20^{\circ }$, the above parameters give rise
to the oscillatory motion of the DD soliton with a period $\sim 10$ ms and
an amplitude $\sim 1$ $\mathrm{\mu }$m.

In summary, we have shown that the interplay of the SOC (spin-orbit
coupling), Raman coupling, and intrinsic nonlinearity in quasi-1D BEC may
realize the mechanism of SOC in the form of mechanical motion of bright and
dark solitons, considered as macroscopic quantum bodies. The soliton's
angular momentum (spin) evolves according to the Bloch equation under the
action of the effective magnetic field, and induces a force affecting the
motion of the soliton's central coordinate. The results have been obtained
by means of the variational analysis and numerical simulations, which
demonstrate a very good agreement. These findings suggest new directions for
experimental studies of the dynamics of matter-wave solitons under the
action of SOC.

\begin{acknowledgments}
We thank Hui Zhai and U. Zuelicke for valuable discussions. This work is
supported by NSFC under grant Nos. 11474205, 11404225, and 11504037. L. Wen
is also supported by Chongqing Research Program of Basic Research and
Frontier Technology under Grant No. cstc2015jcyjA50024, and Foundation of
Education Committees of Chongqing under Grant No. KJ1500311. The work of
B.A.M. is supported, in part, by the joint program in physics between the
National Science Foundation (US) and Binational Science Foundation
(US-Israel), through grant No. 2015616. G.J. was supported by the Lithuanian
Research Council (Grant No. MIP-086/2015).
\end{acknowledgments}


\begin{thebibliography}{99}
\bibitem{Randy} K. E. Strecker, G. B. Partridge, A. G. Truscott, and R. G.
Hulet, New J. Phys. \textbf{5}, 73 (2003).

\bibitem{Brazh} V. A. Brazhnyi and V. V. Konotop, Mod. Phys. Lett. B \textbf{%
18}, 627 (2004).

\bibitem{Fatkh} F. Kh. Abdullaev, A. Gammal, A. M. Kamchatnov, and L. Tomio,
Int. J. Mod. Phys. B \textbf{19}, 3415 (2005).

\bibitem{Morsch} O. Morsch and M. Oberthaler, Rev. Mod. Phys. \textbf{78},
179 (2006).

\bibitem{fsm} B. A. Malomed, \textit{Soliton Management in Periodic Systems}
(Springer, Berlin, 2006).

\bibitem{soliton1} P. G. Kevrekidis, D. J. Frantzeskakis, R. Carretero-Gonz%
\'{a}lez, \textit{Emergent Nonlinear Phenomena in Bose-Einstein Condensates}
(Springer, Berlin, 2008).

\bibitem{Dim} D. J. Frantzeskakis, J. Phys. A: Math. Theor. \textbf{43},
213001 (2010),

\bibitem{RMP2} Y. V. Kartashov, B. A. Malomed, and L. Torner, Rev. Mod.
Phys. \textbf{83}, 247 (2011).

\bibitem{vs} P. G. Kevrekidis and D. J. Frantzeskakis, arXiv:1512.06754.

\bibitem{bs1} S. Burger, K. Bongs, S. Dettmer, W. Ertmer, K. Sengstock, A.
Sanpera, G. V. Shlyapnikov, and M. Lewenstein, Phys. Rev. Lett. \textbf{83},
5198 (1999).

\bibitem{soliton2} L. Khaykovich, F. Schreck, G. Ferrari, T. Bourdel, J.
Cubizolles, L. D. Carr, Y. Castin, C. Salomon, Science \textbf{296}, 1290
(2002).

\bibitem{soliton3} K. E. Strecker, G. B. Partridge, A. G. Truscott, and R.
G. Hulet, Nature (London) \textbf{417}, 150 (2002).

\bibitem{soliton4} S. L. Cornish, S. T. Thompson, and C. E. Wieman, Phys.
Rev. Lett. \textbf{96}, 170401 (2006); A. L. Marchant, T. P. Billam, T. P.
Wiles, M. M. H. Yu, S. Gardiner, and S. L. Cornish, Nature Commun. \textbf{4}%
, 1865 (2013).

\bibitem{soliton5} J. H. V. Nguyen, P. Dyke, D. Luo, B. A. Malomed, and R.
G. Hulet, Nature Phys. \textbf{10}, 918 (2014).

\bibitem{soc1} Y. -J. Lin, K. Jim\'{e}nez-Garc\'{\i}a, and I. B. Spielman,
Nature (London) \textbf{471}, 83 (2011).

\bibitem{soc2} J. -Y. Zhang, S. -C. Ji, Z. Chen, L. Zhang, Z. -D. Du, B.
Yan, G. -S. Pan, B. Zhao, Y. -J. Deng, H. Zhai, S. Chen, and J. -W. Pan,
Phys. Rev. Lett. \textbf{109}, 115301 (2012).

\bibitem{soc3} P. J. Wang, Z. -Q. Yu, Z. K. Fu, J. Miao, L. H. Huang, S. J.
Chai, H. Zhai, and J. Zhang, Phys. Rev. Lett. \textbf{109}, 095301 (2012).

\bibitem{soc4} L. W. Cheuk, A. T. Sommer, Z. Hadzibabic, T. Yefsah, W. S.
Bakr, and M. W. Zwierlein, Phys. Rev. Lett. \textbf{109}, 095302 (2012).

\bibitem{Zhaih} C. J. Wang, C. Gao, C.-M. Jian, and H. Zhai, Phys. Rev.
Lett. \textbf{105}, 160403 (2010); H. Zhai, Int. J. Mod. Phys. B \textbf{26}%
, 1230001 (2012).

\bibitem{TLHo} T.-L. Ho and S. Z. Zhang, Phys. Rev. Lett. \textbf{107},
150403 (2011).

\bibitem{Lyou} Z. F. Xu, R. L\"{u}, and L. You, Phys. Rev. A \textbf{83},
053602 (2011).

\bibitem{CWZhang1} Y. P. Zhang, L. Mao, and C. W. Zhang, Phys. Rev. Lett.
\textbf{108}, 035302 (2012).

\bibitem{YLi} Y. Li, L. P. Pitaevskii, and S. Stringari, Phys. Rev. Lett.
\textbf{108}, 225301 (2012).

\bibitem{SZhang} J. P. Vyasanakere, S. Zhang, and V. B. Shenoy, Phys. Rev. B
\textbf{84}, 014512 (2011).

\bibitem{HHu} H. Hu, L. Jiang, X.-J. Liu, and H. Pu, Phys. Rev. Lett.
\textbf{107}, 195304 (2011).

\bibitem{CWZhang2} M. Gong, S. Tewari, and C. W. Zhang, Phys. Rev. Lett.
\textbf{107}, 195303 (2011).

\bibitem{ZQYu} Z.-Q. Yu and H. Zhai, Phys. Rev. Lett. \textbf{107}, 195305
(2011).

\bibitem{CWZhang3} M. Gong, G. Chen, S.-T. Jia, and C. W. Zhang, Phys. Rev.
Lett. \textbf{109}, 105302 (2012).

\bibitem{CWZhang4} C.-L. Qu, Z. Zheng, M. Gong, et al., Nature Communication
\textbf{4}, 2710 (2013).

\bibitem{WYi} W. Zhang, W. Yi, Nature Communication \textbf{4}, 2711 (2013).

\bibitem{Galitski2013} V. Galitski and I. B. Spielman, Nature \textbf{494}
49 (2013).

\bibitem{Goldman2014} N. Goldman, G. Juzeli\={u}nas, P. \"{O}hberg and I.
B. Spielman, Rep. Prog. Phys. \textbf{77}, 126401 (2014).

\bibitem{Zhai2015} H. Zhai, Rep. Prog. Phys. \textbf{78}, 026001 (2014).

\bibitem{Su2016} S.-W. Su, S.-C. Gou, Q. Sun, L. Wen, W.-M. Liu, A.-C. Ji,
J. Ruseckas, and G. Juzeli\={u}nas, Phys. Rev. A \textbf{93}, 053630 (2016).

\bibitem{ss1} Y. Xu, Y. Zhang, and B. Wu, Phys. Rev. A \textbf{87}, 013614
(2013).

\bibitem{ss4} Y. V. Kartashov, V. V. Konotop, and F. K. Abdullaev, Phys.
Rev. Lett. \textbf{111}, 060402 (2013).

\bibitem{ss3} V. Achilleos, J. Stockhofe, P. G. Kevrekidis, D. J.
Frantzeskakis and P. Schmelcher, Europhys. Lett. \textbf{103}, 20002 (2013).

\bibitem{ss2} V. Achilleos, D. J. Frantzeskakis, P. G. Kevrekidis, and D. E.
Pelinovsky, Phys. Rev. Lett. \textbf{110}, 264101 (2013).

\bibitem{ss5} Y. K. Liu and S. J. Yang, Europhys. Lett. \textbf{108}, 30004
(2014).

\bibitem{ss6} Y. V. Kartashov, V. V. Konotop, and D. A. Zezyulin, Phys. Rev.
A \textbf{90}, 063621 (2014).

\bibitem{ss7} S. Gautam and S. K. Adhikari, Laser Phys. Lett. \textbf{12},
045501 (2015).

\bibitem{ss8} S. Gautam and S. K. Adhikari, Phys. Rev. A \textbf{91}, 063617
(2015).

\bibitem{ss9} Y. Zhang, Y. Xu, and T. Busch, Phys. Rev. A \textbf{91},
043629 (2015)

\bibitem{bdd} V. Achilleos, D. J. Frantzeskakis, and P. G. Kevrekidis, Phys.
Rev. A \textbf{89}, 033636 (2014).

\bibitem{sgs} S. Peotta, F. Mireles, and  M. Di Ventra, Phys. Rev. A \textbf{91}, 021601 (2015).

\bibitem{2cs1} H. Sakaguchi, B. Li, and B. A. Malomed, Phys. Rev. E \textbf{%
89}, 032920 (2014).

\bibitem{Luca} L. Salasnich, W. B. Cardoso, and B. A. Malomed, Phys. Rev. A
\textbf{90}, 033629 (2014).

\bibitem{2cs2} H. Sakaguchi and B. A. Malomed, Phys. Rev. E \textbf{90},
062922 (2014).

\bibitem{hvgs} V. E. Lobanov, Y. V. Kartashov, and V. V. Konotop, Phys. Rev.
Lett. \textbf{112}, 180403 (2014).

\bibitem{Petra} P. Beli\v{c}ev, G. Gligori\'{c}, J. Petrovi\'{c}, A.
Maluckov, L. Hadzievski, and B. Malomed, J. Phys. B\ At. Mol. Opt. Phys.
\textbf{48}, 065301 (2015).

\bibitem{YCZhang} Y.-C. Zhang, Z.-W. Zhou, B. A. Malomed, and H. Pu, Phys.
Rev. Lett. \textbf{115}, 253902 (2015).

\bibitem{Fialko} O. Fialko, J. Brand,and U. Zulicke, Phys. Rev. A \textbf{85}%
, 051605(R) (2012).

\bibitem{RMP1} Y. S. Kivshar, B. A. Malomed, Rev. Mod. Phys. \textbf{61},
763 (1989).

\bibitem{KivPel} Yu. S. Kivshar and D. E. Pelinovsky, Phys. Rep. \textbf{331}%
, 117 (2000).

\bibitem{KA} Y. S. Kivshar and G. P. Agrawal, \textit{Optical Solitons: From
Fibers to Photonic Crystals} (Academic Press, San Diego, 2003).

\bibitem{bullets} B. A. Malomed, D. Mihalache, F. Wise, and L. Torner, J.
Optics B: Quant. Semicl. Opt. \textbf{7}, R53 (2005).

\bibitem{DesTorKiv} A. S. Desyatnikov, L. Torner, and Y. S. Kivshar, Progr.
Opt. \textbf{47}, 1 (2005).

\bibitem{Dum} D. Mihalache, Rom. J. Phys. \textbf{57}, 352 (2012).

\bibitem{PT} V. V. Konotop, J. Yang, and D. A. Zezyulin, Rev. Mod. Phys.
\textbf{88}, 035002 (2016).

\bibitem{d10} T. Busch and J. R. Anglin, Phys. Rev. Lett. \textbf{84}, 2298
(2000); T. Busch and J. R. Anglin, Phys. Rev. Lett. \textbf{87}, 010401
(2001). 

\bibitem{d9} L. D. Carr and Y. Castin, Phys. Rev. A \textbf{66}, 063602
(2002); L. Salasnich, Phys. Rev. A \textbf{70}, 053617 (2004); Z. X. Liang,
Z. D. Zhang, and W. M. Liu, Phys. Rev. Lett. \textbf{94}, 050402 (2005).

\bibitem{d7} L. Li, B. A. Malomed, D. Mihalache, and W. M. Liu, Phys. Rev. E
\textbf{73} 066610 (2006).

\bibitem{d11} A. Weller, J. P. Ronzheimer, C. Gross, J. Esteve, and M. K.
Oberthaler, D. J. Frantzeskakis, G. Theocharis and P. G. Kevrekidis, Phys. Rev. Lett. \textbf{101}, 130401 (2008).

\bibitem{HPu} X. X. Liu, H. Pu, B. Xiong, W. M. Liu, and J. Gong, Phys. Rev.
A \textbf{79}, 013423 (2009).

\bibitem{d12} C. Becker, S. Stellmer, P. Soltan-Panahi, S. D\"{o}rscher, M.
Baumert, E.-M. Richter, J. Kronj\"{a}ger, K. Bongs, and K. Sengstock, Nature
Phys. \textbf{4}, 496 (2008).

\bibitem{ChenZ} Z. Chen and H. Zhai, Phys. Rev. A \textbf{86}, 041604 (2012).

\bibitem{LiY} Y. Li, G. I. Martone, and S. Stringari, Europhys. Lett.
\textbf{99}, 56008 (2012).

\bibitem{manakov} S. V. Manakov, Sov. Phys. JETP \textbf{38}, 248 (1974).

\bibitem{Progress} B. A. Malomed, Progr. Optics \textbf{43}, 71 (2002).

\bibitem{phasemixing} R. Navarro, R. Carretero-Gonz\'{a}lez, and P. G.
Kevrekidis, Phys. Rev. A \textbf{80}, 023613 (2009); L. Wen, W. M. Liu, Y.
Cai, J. M. Zhang, and J. Hu, Phys. Rev. A \textbf{85}, 043602 (2012).

\bibitem{Jacobi} I. S. Gradshteyn and I. M. Ryzhik, \textit{Tables of
integrals, series, and products, Seventh Edition} (Elsevier: Amsterdam,
2007).

\bibitem{r1} I. V. Barashenkov and A. O. Harin, Phys. Rev. Lett. \textbf{72}%
, 1575 (1994).

\bibitem{r2} R. Carretero-Gonz\'{a}lez, D. J. Frantzeskakis, and P. G.
Kevrekidis, Nonlinearity \textbf{21}, 139 (2008).

\bibitem{r3} D. J. Frantzeskakis, J. Phys. A Math. Theor. \textbf{43},
213001 (2010).

\bibitem{r4} Y. S. Kivshar and X. Yang, Phys. Rev. E \textbf{49}, 1657
(1994).

\bibitem{r5} I. M. Uzunov and V. S. Gerdjikov, Phys. Rev. A \textbf{47},
1582 (1993).
\end{thebibliography}
\end{document}


\title{Supplementary Material}
\author{Lin Wen}
\affiliation{Department of Physics, Chongqing Normal University, Chongqing, 401331, China}
\affiliation{Department of Physics, Capital Normal University, Beijing 100048, China}
\author{Q. Sun}
\email{QingSun@cnu.edu.cn}
\affiliation{Department of Physics, Capital Normal University, Beijing 100048, China}
\author{Yu Chen}
\affiliation{Department of Physics, Capital Normal University, Beijing 100048, China}
\author{Deng-Shan Wang}
\affiliation{School of Applied Science, Beijing Information Science and Technology
University, Beijing 100192, China}
\author{J. Hu}
\affiliation{Department of Physics, Capital Normal University, Beijing 100048, China}
\author{H. Chen}
\affiliation{Institute of Physics, Chinese Academy of Sciences, Beijing 100190, China}
\author{W.-M. Liu}
\affiliation{Institute of Physics, Chinese Academy of Sciences, Beijing 100190, China}
\author{G. Juzeli\={u}nas}
\affiliation{Institute of Theoretical Physics and Astronomy, Vilnius University, Saul\.{e}%
tekio Ave. 3, LT-10222 Vilnius, Lithuania}
\author{Boris A. Malomed}
\affiliation{Department of Physical Electronics, School of Electrical Engineering,Faculty
of Engineering, Tel Aviv University, Tel Aviv 69978, Israel}
\affiliation{Laboratory of Nonlinear-Optical Informatics, ITMO University, St. Petersburg
197101, Russia}
\author{An-Chun Ji}
\email{anchun.ji@cnu.edu.cn}
\affiliation{Department of Physics, Capital Normal University, Beijing 100048, China}
\maketitle

In this supplementary material, we present, at first, details of the EL
(Euler-Lagrange) equations. Using the Lagrangian given by Eq. (4) in the
main text, one can derive the standard EL equations as $\partial L/\partial
A-\frac{d}{dt}\left( \partial L/\partial \dot{A}\right) =0$, where $A$
stands for any variational parameter, and $\dot{A}\equiv dA/dt$. Thus, we
derive the following set of evolution equations for the variational
parameters:
\begin{gather}
\frac{dk_{+}}{dt}=\cos \left( 2\theta \right) \frac{dk_{-}}{dt}+\frac{%
2\Omega \pi k_{-}^{2}\sin \left( 2\theta \right) \sin \left( 2\varphi
_{-}-2k_{-}\xi /\eta \right) }{\eta \sinh \left( \pi k_{-}/\eta \right) }\,,
\tag{S1}  \label{s1} \\
\frac{\eta ^{2}}{3}+\frac{1}{6}\eta g=\left[ \frac{\Omega \pi k_{-}}{\eta
\sinh \left( \pi k_{-}/\eta \right) }-\frac{\Omega \pi ^{2}k_{-}^{2}\cosh
\left( \pi k_{-}/\eta \right) }{\eta ^{2}\sinh ^{2}\left( \pi k_{-}/\eta
\right) }\right] \sin \left( 2\theta \right) \cos \left( 2\varphi _{-}-2%
\frac{k_{-}}{\eta }\xi \right) \,,  \tag{S2}  \label{s2} \\
\frac{d\varphi _{-}}{dt}=\frac{\xi }{\eta }\frac{dk_{-}}{dt}%
-k_{+}k_{-}+\lambda k_{+}-\delta -\frac{\Omega \pi k_{-}\cot \left( 2\theta
\right) }{\eta \sinh \left( \pi k_{-}/\eta \right) }\cos \left( 2\varphi
_{-}-2\frac{k_{-}}{\eta }\xi \right) \,,  \tag{S3}  \label{s3} \\
\frac{d\theta }{dt}=-\frac{\Omega \pi k_{-}}{\eta \sinh \left( \pi
k_{-}/\eta \right) }\sin \left( 2\varphi _{-}-2\frac{k_{-}}{\eta }\xi
\right) \,,  \tag{S4}  \label{s4} \\
\frac{d}{dt}\left( \frac{\xi }{\eta }\right) =-k_{+}+\left( k_{-}-\lambda
\right) \cos \left( 2\theta \right) \,,  \tag{S5}  \label{s5} \\
\frac{\eta ^{2}}{3}+\frac{1}{6}\eta g=-k_{-}\left( k_{-}-\lambda \right)
\sin ^{2}\left( 2\theta \right) \,.  \tag{S6}  \label{s6}
\end{gather}%
Here, Eqs. (\ref{s2}) and (\ref{s6}) amount to constant values of the
respective parameters, $\eta =-g/2$ and $k_{-}=\lambda $, under condition $%
\pi \lambda /\eta <<1$. Thus, we arrive at Eqs. (5)-(8) written in the main
text.

Another part of the supplementary material aims to display, in Fig. \ref%
{sup-fig}, results of simulations of the GP system for the initially
polarized state with $\theta (t=0)=\pi /2$ and $\varphi _{-}(t=0)=0$. For $%
\delta =\lambda c_{1}=\lambda ^{2}$, we see that the soliton's spin rotates
periodically on the Bloch sphere, resulting in the oscillations of the
density of the soliton's components.
\begin{figure}[tbp]
\includegraphics[width= 0.48\textwidth]{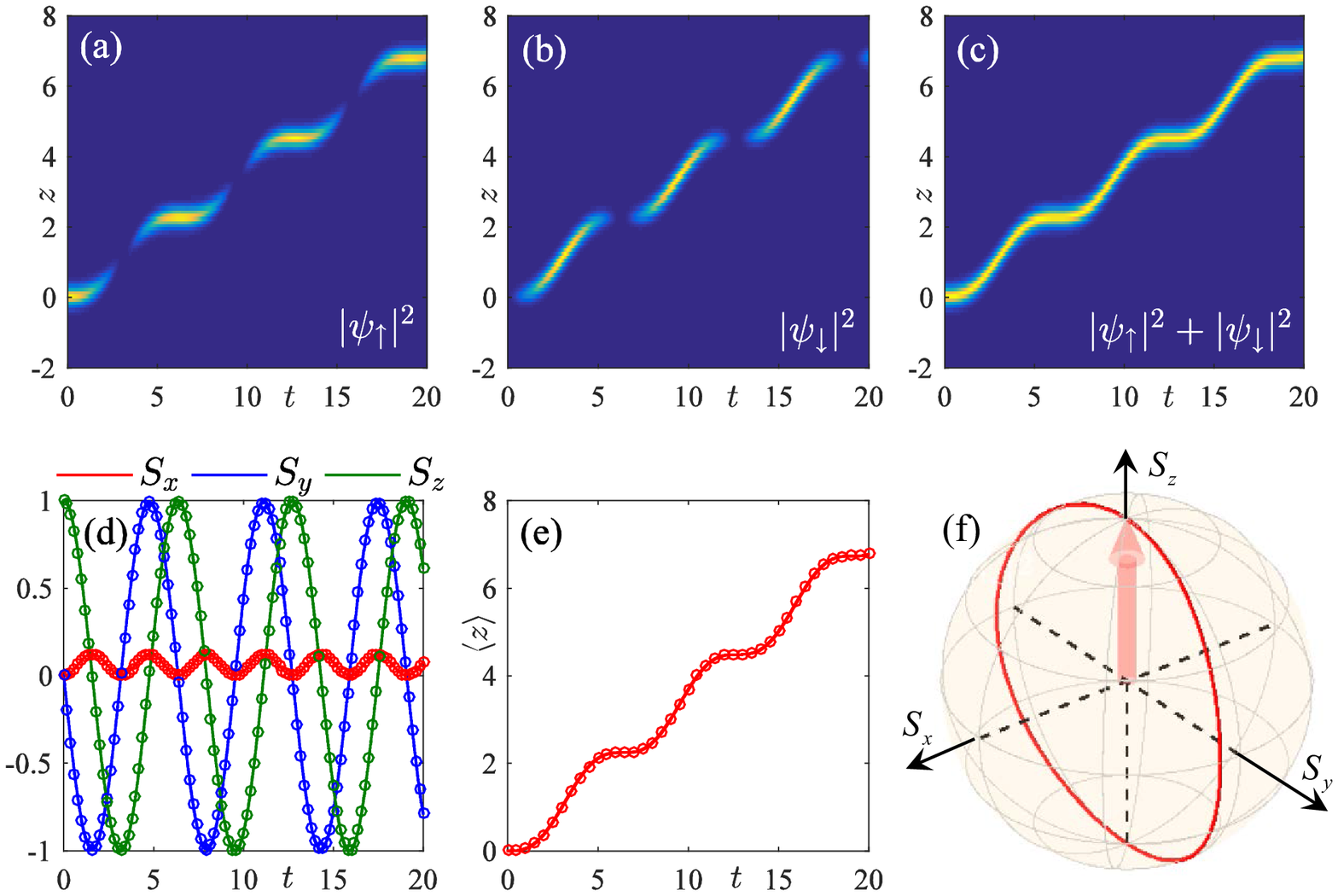}
\caption{(Color online). (a)-(c) The evolution of the density in the two
components of the soliton produced by the GP simulations for the initially
polarized state with $\protect\theta (t=0)=\protect\pi /{2}$ and $\protect%
\varphi _{-}(t=0)=0$. The parameters are $\Omega =0.5$, $\protect\lambda =0.5%
\protect\sqrt{\Omega }$, $\protect\delta =\protect\lambda c_{1}=\protect%
\lambda ^{2}$, and $g=-10$. The spin dynamics and center-of-mass motion of
the soliton, generated by these simulations (circles), and by the
variational approximation (solid lines) are depicted in panels (d)-(e).
Panel (f) displays the track of the spin density on the Bloch sphere.}
\label{sup-fig}
\end{figure}